\documentclass[11pt]{article}
\usepackage{amsmath,amssymb}

\newtheorem{theorem}{Theorem}[section]
\newtheorem{lemma}[theorem]{Lemma}

\newtheorem{proposition}[theorem]{Proposition}
\newtheorem{corollary}[theorem]{Corollary}
\newtheorem{definition}[theorem]{Definition}

\newenvironment{proof}{{\bf Proof:\ }}{\hfill$\Box$\medskip}

\newcommand{\ignore}[1]{}

\newcommand\npf{\mbox{ }\hfill\sqr\vskip6pt}
\def\sqr{$\vcenter{\hrule height.2mm
\hbox{\vrule width.2mm height2mm\kern2mm \vrule width.2mm}\hrule height.2mm}$}

\def \R {{\mathbb R}}
\def \Z {{\mathbb Z}}

\begin{document}

\title{
Finding heaviest $H$-subgraphs in real weighted graphs, with applications
\thanks
{This paper is based upon combining two preliminary versions \cite{VW06,VWY06} appearing in
{\em Proceedings of the 38th Annual ACM Symposium on Theory of Computing (STOC)}, Seattle, WA, 2006, and
{\em Proceedings of the 33rd International Colloquium on Automata, Languages and Programming (ICALP)}, Venice, Italy 2006.
}
}

\author{
Virginia Vassilevska
\thanks{
Computer Science Department, Carnegie Mellon University, Pittsburgh, PA. email: virgi@cs.cmu.edu}
\and
Ryan Williams
\thanks{
Computer Science Department, Carnegie Mellon University, Pittsburgh, PA. email: ryanw@cs.cmu.edu}
\and
Raphael Yuster
\thanks{
Department of Mathematics, University of Haifa, Haifa, Israel. email: raphy@math.haifa.ac.il}
}

\date{}

\maketitle

\begin{abstract}
For a graph $G$ with real weights assigned to the vertices (edges),
the MAX $H$-SUBGRAPH problem is to find an $H$-subgraph of $G$ with maximum total weight,
if one exists. Our main results are new strongly polynomial algorithms for the MAX $H$-SUBGRAPH problem.
Some of our algorithms are based, in part, on fast matrix multiplication.

For vertex-weighted graphs with $n$ vertices we solve a more general problem -- the \emph{all pairs} MAX $H$-SUBGRAPH problem, where the task is to find for every pair of vertices $u,v$, a maximum $H$-subgraph containing both $u$ and $v$, if one exists.
We obtain 
an $O(n^{t(\omega,h)})$ time
algorithm for the \emph{all pairs} MAX $H$-SUBGRAPH problem in the case where $H$ is a fixed
graph with $h$ vertices and $\omega < 2.376$ is the exponent of
matrix multiplication. The value of $t(\omega,h)$ is determined by
solving a small integer program. In particular, heaviest
triangles for all pairs can be found in $O(n^{2+1/(4-\omega)}) \le o(n^{2.616})$
time. 
For
$h=4,5,8$ the running time of our algorithm essentially matches
that of the (unweighted) $H$-subgraph detection problem.
Using rectangular matrix multiplication, the value of
$t(\omega,h)$ can be improved; for example, the runtime for
triangles becomes $O(n^{2.575})$.

We also present improved algorithms for the MAX $H$-SUBGRAPH problem in the edge-weighted case.
In particular, we obtain an $O(m^{2-1/k} \log n)$ time algorithm for the heaviest cycle
of length $2k$ or $2k-1$ in a graph with $m$ edges and an $O(n^3/\log n)$ time randomized algorithm for finding the heaviest cycle of any fixed length.

Our methods also yield efficient algorithms for several related problems that are faster than any previously existing algorithms. For example, we show how to find chromatic $H$-subgraphs in edge-colored graphs, and how to compute the most significant bits of the distance product of two real matrices, in truly sub-cubic time.


\end{abstract}

\section{Introduction}
Finding cliques or other types of subgraphs in a larger graph
are classical problems in complexity theory and algorithmic combinatorics.
Finding a maximum clique is NP-Hard, and also hard to approximate \cite{Ha98}.
This problem is also conjectured to be {\em not} fixed parameter tractable \cite{DoFe95}.
The problem of finding (induced) subgraphs on $k$ vertices in an $n$-vertex graph has
been studied extensively
(see, e.g., \cite{AlYuZw95,AlYuZw97,ChNi85,EiGr04,KlKrMu00,NePo85,PaYa81,YuZw04}).
All known algorithms for finding an induced subgraph on $k$ vertices have running time $n^{\Theta(k)}$. Many of these algorithms use fast matrix multiplication to obtain
improved exponents.

The main contribution of this paper is a set of improved
algorithms for finding (induced) $k$-vertex subgraphs in a real
vertex-weighted or edge-weighted graph. More formally, let $G$ be
a graph with real weights assigned to the vertices (edges). The
weight of a subgraph of $G$ is the sum of the weights of its
vertices (edges). The MAX $H$-SUBGRAPH problem is to find an
$H$-subgraph of maximum weight, if one exists. Some of our
algorithms are based, in part, on {\em fast} matrix
multiplication. In several cases, our algorithms use fast {\em
rectangular\/} matrix multiplication algorithms.
However, for simplicity reasons, we express most of our time
bounds in terms of $\omega$, the exponent of fast {\em square\/}
matrix multiplications. The best bound currently available on
$\omega$ is $\omega < 2.376$, obtained by Coppersmith and Winograd
\cite{CoWi90}. This is done by reducing each rectangular matrix
product to a collection of smaller square matrix products.
Slightly improved bounds can be obtained by using the best
available rectangular matrix multiplication algorithms of
Coppersmith \cite{Co97} and Huang and Pan \cite{HuPa98}. In all of
our algorithms we assume that the graphs are {\em undirected}, for
simplicity. All of our results are applicable to directed graphs
as well. Likewise, all of our results on the MAX $H$-SUBGRAPH
problem hold for the analogous MIN $H$-SUBGRAPH problem. As usual,
we use the {\em addition-comparison} model for handling real
numbers. That is, real numbers are only allowed to be compared or
added. In particular, our algorithms are strongly polynomial.

Our first algorithm applies to {\em vertex-weighted} graphs.
In order to describe its complexity we need to define a small (constant size) integer optimization problem.
Let $h \ge 3$ be a positive integer. The function $t(\omega,h)$ is defined by the following
optimization program.
\begin{definition}
\begin{equation}\label{e1}
b_1 = \max \{b \in N ~:~ \frac{b}{4-\omega} ~\le~ \lfloor \frac{h-b}{2} \rfloor\}.
\end{equation}
\begin{equation}\label{e2}
s_1 = h-b_1+\frac{b_1}{4-\omega}.
\end{equation}
\begin{equation}\label{e3}
s_2(b) = \max \{h-b+\lfloor \frac{h-b}{2} \rfloor ~,~h-(3-\omega)\lfloor \frac{h-b}{2} \rfloor \}.
\end{equation}
\begin{equation}\label{e4}
s_2 = \min \{s_2(b) ~:~ \lfloor \frac{h-b}{2} \rfloor \le b \le h-2\}.
\end{equation}
\begin{equation}\label{e5}
t(\omega,h) = \min \{s_1,s_2\}.
\end{equation}
\end{definition}
By using fast rectangular matrix multiplication, an alternative definition for
$t(\omega,h)$, resulting in slightly smaller values, can be obtained
(note that if $\omega=2$, as conjectured by many researchers, fast rectangular matrix multiplication has no advantage over fast square matrix multiplication).
\begin{theorem}
\label{t-vertex-real}
Let $H$ be a fixed graph with $h$ vertices.
Let $G=(V,E)$ be a graph with $n$ vertices, and $w : V \rightarrow \Re$ a weight function.
For every pair of vertices $u,v \in V$, an induced $H$-subgraph of $G$ containing $u$ and $v$ of maximum weight
(if one exists), can be found in $O(n^{t(\omega,h)})$ time. In particular, the MAX $H$-subgraph problem can be solved in
$O(n^{t(\omega,h)})$ time.
\end{theorem}
Notice that Theorem \ref{t-vertex-real} solves, in fact, a more general problem, the {\em All-Pairs} MAX $H$-Subgraph problem.
It is easy to establish some small values of $t(\omega,h)$
directly. For $h=3$ we have $t(\omega,3)=2+1/(4-\omega) < 2.616$
by taking $b_1=1$ in (\ref{e1}). Using fast rectangular matrix
multiplication this can be improved to $2.575$. In particular, for each pair of vertices, a
triangle of maximum weight containing them (if one exists) can be found in $o(n^{2.575})$ time.
This should be compared to the well-known $O(n^\omega) \le o(n^{2.376})$ time
algorithm for detecting a triangle in an {\em unweighted} graph \cite{ItRo78}.
For $h=4$ we have $t(\omega,4)=\omega+1 < 3.376$ by taking $b=2$
in (\ref{e4}). Interestingly, the fastest algorithm for detecting
a $K_4$, that uses square matrix multiplication, also runs in
$O(n^{\omega+1})$ time \cite{NePo85}. The same phenomena also
occurs for $h=5$ where $t(\omega,5)=\omega+2 < 4.376$ and for
$h=8$ where $t(\omega,8)=2\omega+2 < 6.752$. We also note that $t(\omega,6)=4+ 2/(4-\omega)$,
$t(\omega,7)=4+3/(4-\omega)$, $t(\omega,9)=2\omega+3$ and
$t(\omega,10)=6+4/(4-\omega)$. However, a closed formula for
$t(\omega,h)$ cannot be given. Already for $h=11$, and for
infinitely many values thereafter, $t(\omega,h)$ is only piecewise
linear in $\omega$. For example, if $7/3 \le \omega < 2.376$ then
$t(\omega,11)=3\omega+2$, and if $2 \le \omega \le 7/3$ then
$t(\omega,11)=6+5/(4-\omega)$. Finally, it is easy to verify that
both $s_1$ in (\ref{e2}) and $s_2$ in (\ref{e4}) converge to
$3h/(6-\omega)$ as $h$ increases. Thus, $t(\omega,h)$ converges to
$3h/(6-\omega) < 0.828h$ as $h$ increases.

Prior to this work, the only known algorithm for
MAX $H$-SUBGRAPH in the vertex-weighted case (moreover, the All-Pairs version of the problem) was the
na\"{\i}ve $O(n^h)$ algorithm. In
general, reductions to fast matrix multiplication tend to fail
miserably in the case of real-weighted graph problems. The most
prominent example of this is the famous All-Pairs Shortest Paths
(APSP) problem. Seidel \cite{Se95} and Galil and
Margalit~\cite{GaMa97} developed $\tilde{O}(n^{\omega})$ algorithms
for undirected unweighted graphs. However, for arbitrary
edge weights, the best published algorithm known is a recent $O(n^3/\log
n)$ by Chan \cite{Ch05}. When the edge weights are integers in
$[-M,M]$, the problem is solvable in $\tilde{O}(Mn^\omega)$ by
Shoshan and Zwick \cite{ShZw99}, and $\tilde{O}(M^{0.681}n^{2.575})$ by Zwick \cite{Zw02},
respectively. Earlier, a series of papers in the 70's and 80's
starting with Yuval \cite{Yu76} attempted to speed up APSP
directly using fast matrix multiplication. Unfortunately, these
works require a model that allows infinite-precision operations in
constant time.

A slight modification in the algorithm of Theorem \ref{t-vertex-real}, without increasing its
running time by more than a logarithmic factor,
can also answer the decision problem: ``for every pair of vertices $u,v$, is there an $H$-subgraph containing $u$ and $v$, whose weight is
in the interval $[w_1,w_2]$ where $w_1 \le w_2$ are two given reals?''
Another feature of Theorem \ref{t-vertex-real} is that it makes a relatively small number of
comparisons. For example, a heaviest triangle can be found by the algorithm using only $O(m+n \log n)$ comparisons, where $m$ is the number of edges of $G$.

Since Theorem \ref{t-vertex-real} is stated for induced $H$-subgraphs, it obviously also applies to
not-necessarily induced $H$-subgraphs. However, the latter problem can, in some cases, be solved faster.
For example, we show that the $o(n^{2.616})$ time bound for finding a heaviest triangle also holds if one searches for a heaviest $H$-subgraph in the case when $H$ is the complete bipartite graph $K_{2,k}$.

Several $H$-subgraph detection algorithms take advantage of the
fact that $G$ may be sparse. Improving a result of Itai and Rodeh
\cite{ItRo78}, Alon, Yuster and Zwick obtained an algorithm for
detecting a triangle, expressed in terms of $m$ \cite{AlYuZw97}.
The running time of their algorithm is $O(m^{2\omega/(\omega+1)})
\le o(m^{1.41})$. This is faster than the $O(n^\omega)$ algorithm
when $m = o(n^{(\omega+1)/2})$. The best known running times in
terms of $m$ for $H=K_k$ when $k \ge 4$ are given in
\cite{EiGr04}. Sparseness can also be used to obtain faster
algorithms for the vertex-weighted MAX $H$-SUBGRAPH problem. We prove:
\begin{theorem}
\label{t-vertex-m}
If $G=(V,E)$ is a graph with $m$ edges and no isolated vertices, and $w : V \rightarrow \Re$ is a weight function, then a triangle of $G$ with maximum weight (if one exists) can be found in
$O(m^{(18-4\omega)/(13-3\omega)}) \le o(m^{1.45})$ time.
\end{theorem}

The proofs of Theorems \ref{t-vertex-real} and \ref{t-vertex-m}, and some of their consequences, appear in Section 3.
In Section 2 we first introduce a general method called {\em dominance computation}, motivated by a problem in computational geometry and introduced by Matousek in \cite{Ma91}, and show how it can be used to obtain a truly sub-cubic
algorithm for the MAX $K_3$-SUBGRAPH problem. Although the running time we obtain using this method is slightly inferior to that of Theorem \ref{t-vertex-real}, we show that this method has other very interesting applications.
In fact, we will show how to use it in order to efficiently solve a general buyer-seller problem from
computational economics. Another interesting application of the method is the ability to compute the
{\em the most significant bits} of the distance product $A \star B$ of two real matrices, in truly sub-cubic time
(see the definition of distance products in the next section).
Computing the distance product quickly has long been considered as the key to a truly sub-cubic APSP algorithm, since it is known that the time complexity of APSP is no worse than that of the
distance product of two arbitrary $n \times n$ matrices.

We now turn to edge-weighted graphs.
An $O(m^{2-1/\lceil k/2 \rceil})$ time algorithm for detecting the existence of a cycle of length $k$
is given in \cite{AlYuZw97}. A small improvement was obtained later in \cite{YuZw04}.
However, the algorithms in both papers fail when applied to edge-weighted graphs.
Using the {\em color coding} method, together with several additional ideas, we obtain a
randomized $O(m^{2-1/\lceil k/2 \rceil})$ time algorithm in the edge-weighted case, and an
$O(m^{2-1/\lceil k/2 \rceil}\log n)$ deterministic algorithm.
\begin{theorem}
\label{t-edge-cyc}
Let $k \ge 3$ be a fixed integer.
If $G=(V,E)$ is a graph with $m$ edges and no isolated vertices, and $w : E \rightarrow \Re$ is a weight function, then a maximum weight cycle of length $k$, if one exists, can be found with high probability in $O(m^{2-1/\lceil k/2 \rceil})$ time, and deterministically in $O(m^{2-1/\lceil k/2 \rceil}\log n)$ time.
\end{theorem}

In a recent result of Chan \cite{Ch05} it is shown that the distance product
of two $n \times n$ matrices with real entries can be computed in $O(n^3/\log n)$
time (again, reals are only allowed to be compared or added; more recently, Y. Han announced an
$O(n^3(\log \log n/\log n)^{5/4})$ time algorithm).
We show how to reduce the MAX $H$-SUBGRAPH problem in
edge-weighted graphs to the problem of computing a distance
product.
\begin{theorem}
\label{t-edge-real}
Let $H$ be a fixed graph with $h$ vertices.
If $G=(V,E)$ is a graph with $n$ vertices, and $w : E \rightarrow \Re$ is a weight function,
then an induced $H$-subgraph of $G$ (if one exists) of maximum weight can be found in
$O(n^h/\log n)$ time.
\end{theorem}

We can strengthen the above result considerably, in the case where
$H$ is a cycle. For (not-necessarily induced) cycles of fixed
length we can combine distance products with the color coding
method and obtain:
\begin{theorem}
\label{t-edge-cyc-2}
Let $k$ be a fixed positive integer.
If $G=(V,E)$ is a graph with $n$ vertices, and $w : E \rightarrow \Re$ is a weight function,
a maximum weight cycle with $k$ vertices
(if exist) can be found, with high probability, in $O(n^3/\log n)$ time.
\end{theorem}
In fact, the proof of Theorem \ref{t-edge-cyc-2} shows that a maximum weight
cycle with $k=o(\log \log n)$ vertices can be found in (randomized) sub-cubic time.
Section 4 considers edge-weighted graphs and contains the algorithms proving Theorems \ref{t-edge-cyc}, \ref{t-edge-real} and \ref{t-edge-cyc-2}.

Finally, we consider the related problem of finding a certain chromatic $H$-subgraph in an edge-colored graph.
We consider the two extremal chromatic cases. An $H$-subgraph of an edge-colored graph is called {\em rainbow} if all the edges have
distinct colors. It is called {\em monochromatic} if all the edges have the same color. Many combinatorial problems are concerned with the existence of rainbow and/or monochromatic subgraphs.

We obtain a new algorithm that finds a rainbow $H$-subgraph, if one exists.
\begin{theorem}
\label{t-rainbow}
Let $H$ be a fixed graph with $3k+j$ vertices, $j \in \{0,1,2\}$.
If $G=(V,E)$ is a graph with $n$ vertices, and $c : E \rightarrow C$ is
an edge-coloring,
then a rainbow $H$-subgraph of $G$ (if one exists)
can be found in $O(n^{\omega k +j}\log n)$ time.
\end{theorem}
The running time in Theorem \ref{t-rainbow} matches, up to a logarithmic factor, the running time of the induced $H$-subgraph detection problem in (uncolored) graphs.

We obtain a new algorithm that finds a monochromatic $H$-subgraph, if one exists.
For fixed $H$, the running time of our algorithm matches the running time of the
(uncolored) $H$-subgraph detection problem, except for the case $H=K_3$.
\begin{theorem}
\label{t-monochromatic}
Let $H$ be a fixed connected graph with $3k+j$ vertices, $j \in \{0,1,2\}$.
If $G=(V,E)$ is a graph with $n$ vertices, and $c : E \rightarrow C$ is
an edge-coloring,
then a monochromatic $H$-subgraph of $G$ (if one exists) can be found in $O(n^{\omega k +j})$ time, unless $H=K_3$. A monochromatic triangle can be found in
$O(n^{(3+\omega)/2}) \le o(n^{2.688})$ time.
\end{theorem}
The algorithms for edge-colored graphs yielding Theorems
\ref{t-rainbow} and \ref{t-monochromatic} appear in Section 5.
The final section contains some concluding remarks and open problems.

\section{Dominance computations}

Given a set of points $\{v_1,\ldots,v_n\}$ in ${\mathbb R}^d$, the
{\em dominating pairs problem} is to find all pairs of points
$(v_i,v_j)$ such that for all $k=1,\ldots,d$, $v_i[k] \leq
v_j[k]$. The key insight to our method is a connection between the
problem of finding triangles and the well-known problem of
computing dominating pairs in computational geometry. This
connection was inspired by recent work of Chan \cite{Ch05}, who
demonstrated how a $O(c^d n^{1+\varepsilon} + n^2)$ algorithm for
computing dominating pairs in $d$ dimensions can be used to solve
the arbitrary APSP problem in $O(n^3/\log n)$ time.

In particular, we use an elegant algorithm by Matousek for
computing dominating pairs in $n$ dimensions \cite{Ma91}.
Matousek's algorithm does a bit more than determine dominances ---
it actually computes a matrix $D$ such that
\[D[i,j] = |\{k~|~v_i[k] \leq v_j[k]\}|.\] We will call $D$ the
{\em dominance matrix} in the following.

\begin{theorem}[Matousek~\cite{Ma91}] \label{thm:matousek} Given a set $S$ of $n$ points in $\R^n$, the
dominance matrix for $S$ can be computed in
$O\left(n^{\frac{3+\omega}{2}}\right)$ time.
\end{theorem} We outline Matousek's approach
in the following paragraphs. For each coordinate $j=1,\ldots,n$,
sort the $n$ points by coordinate $j$. This takes $O(n^2 \log n)$
time. Define the {\em $j$th rank of point $v_i$}, denoted as
$r_j(v_i)$, to be the position of $v_i$ in the sorted list for
coordinate $j$.

For a parameter $s \in [\log n, n]$, make $n/s$ pairs of Boolean
matrices $(A_1,B_1)\ldots,(A_{n/s},B_{n/s})$ defined as
follows:\[A_k[i,j] = 1 \iff r_j(v_i) \in [ks,ks+s),\]\[B_k[i,j] =
1 \iff r_j(v_i) \geq ks+s.\]

Now, multiply $A_k$ with $B_k^T$, obtaining a matrix $C_k$. Then
$C_k[i,j]$ equals the number of coordinates $c$ such that $v_i[c]
\leq v_j[c]$, $r_c(v_i) \in [ks,ks+s)$, and $r_j(v_i) \geq ks+s$.

Therefore, letting \[C = \sum_{k=1}^{n/s} C_k,\] we have that
$C[i,j]$ is the number of coordinates $c$ such that $\lfloor
r_c(v_i)/s \rfloor < \lfloor r_c(v_j)/s\rfloor$.

Suppose we compute a matrix $E$ such that $E[i,j]$ is the number
of $c$ such that $v_i[c] \leq v_j[c]$ and $\lfloor r_c(v_i)/s
\rfloor = \lfloor r_c(v_j)/s\rfloor$. Then, defining $D := C + E$,
we have the desired matrix \[D[i,j] = |\{k~|~v_i[k] \leq
v_j[k]\}|.\]

To compute $E$, we use the $n$ sorted lists. For an integer $n$, define $[n]= \{1,\ldots ,n\}$. 
Then, for each pair $(i,j)\in [n] \times [n]$, we look up $v_i$'s position $p$ in the sorted
list for coordinate $j$. By reading off the adjacent points less
than $v_i$ in this sorted list ({\em i.e.} the points at positions
$p-1$, $p-2$, {\em etc.}), and stopping when we reach a point
$v_k$ such that $\lfloor r_j(v_k)/s \rfloor < \lfloor
r_j(v_i)/s\rfloor$, we obtain the list $v_{i_1},\ldots,v_{i_\ell}$
of $\ell \leq s$ points such that, for all $x=1,\ldots, \ell$, $v_{i_x}[j]\leq v_i[j]$ and
$\lfloor r_j(v_i)/s \rfloor = \lfloor r_j(v_{i_x})/s\rfloor$. Finally, for each $x
= 1,\ldots,\ell$, we add a 1 to $E[i_x,i]$. Assuming constant time
lookups and constant time probes into a matrix, this entire
process takes only $O(n^2 s)$ time.

The running time of the above procedure is $O(n^2 s + \frac{n}{s}
n^{\omega})$. Choosing $s = n^{\frac{\omega-1}{2}}$, the time bound
becomes $O\left(n^{\frac{3+\omega}{2}}\right)$.

\subsection{Finding a heaviest triangle in sub-cubic time}

We first present a weakly polynomial deterministic
algorithm, then a randomized strongly polynomial algorithm.

\begin{theorem}
\label{det}
On graphs with integer weights, a maximum
vertex-weighted triangle can be found in
$O(n^{(\omega+3)/2}\cdot \log W)~\text{time}$, where $W$ is the maximum weight of a
triangle. On graphs with real weights, a maximum vertex-weighted
triangle can be found in $O(n^{(\omega+3)/2}\cdot B)$ time, where
$B$ is the maximum number of bits in a weight.
\end{theorem}
\begin{proof}
The idea is to obtain a procedure that, given a parameter $K$,
returns an edge $(i,j)$ from a triangle of weight at least $K$.
Then one can binary search to find the weight of the maximum
triangle, and try all possible vertices $k$ to get the triangle
itself.

We first explain the binary search. Without loss of generality, we
assume that all edge weights are at least 1. Let $W$ be the
maximum weight of a triangle. Start by checking if there is a
triangle of weight at least $K=1$ (if not, there are no
triangles). Then try $K=2^i$ for increasing $i$, until there
exists a triangle of weight $2^i$ but no triangle of weight
$2^{i+1}$. This $i$ will be found in $O(\log W)$ steps. After
this, we search on the interval $[2^i,2^{i+1})$ for the largest
$K$ such that there is a triangle of weight $K$. This takes
$O(\log W)$ time for integer weights, and $O(B)$ time for real
weights with $B$ bits of precision.

We now show how to return an edge from a triangle of weight at
least $K$, for some given $K$. Let $V = \{1,\ldots,n\}$ be the set
of vertices. For every $i\in V$, we make a point $f_i =
(e(1),\ldots,e(n))$, where
\[e(j) = \begin{cases} K - w(i) &\text{if there is an edge from
$i$ to $j$}\\ \infty &\text{otherwise.}\end{cases}\] (In
implementation, we can of course substitute a sufficiently large
value in place of $\infty$.) We also make a point $g_i =
(e'(1),\ldots,e'(n))$, where
\[e'(j) =
\begin{cases} w(i)+w(j) & \text{if there is an edge from $i$ to
$j$}\\ -\infty & \text{otherwise.}\end{cases}\] Compute the
dominance matrix $D(K)$ on the sets $\{f_i\}$ and $\{g_i\}$. For
all edges $(i, j)$ in the graph, check if there exists a $k$ such
that $f_i[k] \leq g_j[k]$. This can be done by examining the
appropriate entry in $D(K)$. If such a $k$ exists, then we know
there is a vertex $k$ such that
\[K - w(i) \leq w(j)+w(k) ~\Longrightarrow~ K \leq
w(i)+w(k)+w(j),\] that is, there exists a triangle of weight at
least $K$ using edge $(i, j)$.
Observe that the above works for both directed and undirected
graphs.
\end{proof}

In the above, the binary search over all possible weights prevents
our algorithm from being strongly polynomial. We would like to
have an algorithm that, in a comparison-based model, has a runtime
with {\em no} dependence on the bit lengths of weights. Here we
present a randomized algorithm that achieves this.

\begin{theorem}
\label{rand}
On graphs with real weights, a maximum
vertex-weighted triangle can be found in 
$O(n^{(\omega+3)/2}\cdot \log n)$ expected worst-case time.
\end{theorem}
We would like to somehow binary search over the collection of
triangles in the graph to find the maximum. As this collection is
$O(n^3)$, we would then have our strongly polynomial bound.
Ideally, one would like to pick the ``median'' triangle from a
list of all triangles, sorted by weight. But as the number of
triangles can be $\Omega(n^3)$, forming this list is hopeless.
Instead, we shall show how dominance computations allow us to
efficiently and uniformly sample a triangle at random, whose
weight is from any prescribed interval $(W_1,W_2)$. If we pick a
triangle at random and measure its weight, there is a good chance
that this weight is close to the median weight. In fact, a binary
search that randomly samples for a pivot can be expected to
terminate in $O(\log n)$ time.

Let $W_1, W_2 \in \R \cup \{-\infty,\infty\}$, $W_1 < W_2$,  and
$G$ be a vertex-weighted graph.

\begin{definition} ${\cal C}(W_1,W_2)$
is defined to be the collection of triangles in $G$ whose total
weight falls in the range $[W_1,W_2]$.
\end{definition}

\begin{lemma}\label{sample} One can sample a triangle uniformly at
random from ${\cal C}(W_1,W_2)$, in $O(n^{(\omega+3)/2})$ time.
\end{lemma}
\begin{proof} From the proof of Theorem \ref{det}, one can
compute a matrix $D(K)$ in $O(n^{(\omega+3)/2})$ time, such that
$D(K)[i,j] \neq 0$ iff there is a vertex $k$ such that $(i,k)$ and
$(k,j)$ are edges, and $w(i)+w(j)+w(k) > K$. In fact, the $i,j$
entry of $D(K)$ is the {\em number} of distinct vertices $k$ with
this property. 

Similarly, one can compute matrices $E(K)$ and $L(K)$ such that $E(K)[i,j]$ and 
$L(K)[i,j]$ contain the number of vertices $k$ such that $(i,k)$ and
$(k,j)$ are edges, and $w(i)+w(j)+w(k) \leq K$ (for $E$) or $w(i)+w(j)+w(k) < K$ (for $L$).
(This can be
done by flipping the signs on all coordinates in the sets of
points $\{f_i\}$ and $\{g_i\}$ from Theorem \ref{det}, then
computing dominances, disallowing equalities for $L$.)

Therefore, if we take $F = E(W_2) - L(W_1)$, then $F[i,j]$ is the
number of vertices $k$ where there is a path from $i$ to $k$ to $j$,
and $w(i)+w(j)+w(k) \in [W_1,W_2]$.

Let $f$ be the sum of all entries $F[i,j]$. For each $(i,j) \in
E$, choose $(i,j)$ with probability $F[i,j]/f$. By the above, this
step uniformly samples an edge from a random triangle. Finally, we
look at the set of vertices $S$ that are neighbors to both $i$ and
$j$, and pick each vertex in $S$ with probability $\frac{1}{|S|}$.
This step uniformly samples a triangle with edge $(i,j)$. The
final triangle is therefore chosen uniformly at random.
\end{proof}

\noindent
Observe that there is an interesting corollary to the above.
\begin{corollary}
In any graph, one can sample a triangle uniformly at
random in $O(n^{\omega})$ time.
\end{corollary}
\begin{proof} (Sketch) Multiplying the adjacency matrix with itself counts the
number of 2-paths from each vertex to another vertex. Therefore one
can count the number of triangles and sample just as in the above.
\end{proof}

We are now prepared to give the strongly polynomial algorithm.

\noindent
{\bf Proof of Theorem \ref{rand}:}\,
Start by
choosing a triangle $t$ uniformly at random from all triangles. By
the corollary, this is done in $O(n^{\omega})$ time.

Measure the weight $W$ of $t$. Determine if there is a triangle
with weight in the range $(W,\infty)$, in $O(n^{(\omega+3)/2})$
time. If not, return $t$. If so, randomly sample a triangle from
$(W,\infty)$, let $W'$ be its weight, and repeat the search with
$W'$.

It is routine to estimate the runtime of this procedure, but we
include it for completeness. Let $T(n,k)$ be the expected runtime
for an $n$ vertex graph, where $k$ is the number of triangles in the
current weight range under inspection. In the worst case,
\[T(n,k) \leq \frac{1}{k} \sum_{i=1}^{k-1} T(n,k-i) +
c \cdot n^{(\omega+3)/2}\] for some constant $c \geq 1$. But this
means
\[T(n,k-1) \leq \frac{1}{k-1} \sum_{i=1}^{k-2} T(n,k-i) +
c \cdot n^{(\omega+3)/2},\] so \begin{eqnarray*} T(n,k) & \leq &
\left(\frac{1}{k} + \frac{k-1}{k}\right) \cdot T(n,k-1)\\ && +
\left(1-\frac{k-1}{k}\right)c n^{(\omega+3)/2}\\
& = & T(n,k-1) + \frac{c}{k}n^{(\omega+3)/2},\end{eqnarray*} which
solves to $T(n,k) = O(n^{(\omega+3)/2} \log k)$.\npf

\subsection{Most significant bits of a distance product}

Let $A$ and $B$ be two $n \times n$ matrices with
entries in $\Re \cup \infty$.
The {\em distance product} $C = A \star B$ is an $n \times n$ matrix
with $C[i,j] = \min_{k=1 \ldots ,n} A[i,k]+B[k,j]$.
Clearly, $C$ can be computed in $O(n^3)$ time in the addition-comparison model.
In fact, Kerr \cite{Ke70} showed that the distance product
requires $\Omega(n^3)$ on a straight-line program using $+$ and $\min$. 
However, Fredman showed in \cite{Fr76} that the distance product of
two square matrices of order $n$ can be performed in
$O(n^3(\log \log n/\log n)^{1/3})$ time.
Following a sequence of improvements over Fredman's result, Chan
gave an $O(n^3/\log n)$ time algorithm for distance products.

Computing the distance product quickly has long been considered as
the key to a truly sub-cubic APSP algorithm, since it is known
that the time complexity of APSP is no worse than that of the
distance product of two arbitrary $n \times n$ matrices.
Practically all APSP algorithms with runtime of the form
$O(n^{\alpha})$ have, at their core, some form of distance
product. Therefore, any improvement on the complexity of distance
product is interesting.

Here we show that {\em the most significant bits of $A \star B$}
can be computed in sub-cubic time, again with no exponential
dependence on edge weights. In previous work, Zwick~\cite{Zw02}
shows how to compute {\em approximate} distance products. Given
any $\varepsilon > 0$, his algorithm computes distances $d_{ij}$
such that the difference of $d_{ij}$ and the exact value of the
distance product entry is at most $O(\varepsilon)$. The running
time of his algorithm is $O(\frac{W}{\varepsilon} \cdot n^{\omega}
\log W)$. Unfortunately, guaranteeing that the distances are
within $\varepsilon$ of the right values, does not necessarily
give any of the bits of the distances. Our strategy is to use
dominance matrix computations.

\begin{proposition}
Let $A,B \in (\Z \cup \{+\infty,-\infty\})^{n \times n}$.
The $k$ most significant bits of all entries in $A \star B$ can be
determined in $O(2^k \cdot n^{\frac{3+\omega}{2}} \log n)$ time,
assuming a comparison-based model.
\end{proposition}
\begin{proof}
For a matrix $M$, let $M[i,:]$ be the $i$th row, and $M[:,j]$ be the $j$th
column. For a constant $K$, define the set of vectors \[M^L(K) :=
\{(M[i,1]-K, \ldots, M[i,n]-K)~|~ i=1,\ldots,n\}.\] Also, define
\[M^R(K) := \{(-M[1,i], \ldots, -M[n,i]) ~|~ i=1,\ldots,n\}.\] 

Now
consider the set of vectors $S(K)=A^L(K) \cup B^R(K)$. 
Using a dominance computation on $S(K)$, one can obtain
 the matrix $C(K)$ defined by 
 
\begin{displaymath}
C(K)[i,j] := \left\{\begin{array}{ll}
0 & \textrm{if } \exists k~~ \textrm{s.t. } u_i[k] < v_j[k], u_i\in A^L(K),v_j \in B^R(K)\\
1 & \textrm{ otherwise}
\end{array}\right.
\end{displaymath}
 
%
Then for any
$i,j$, \[\min_k\{A[i,k]+B[k,j]\} \geq K \iff C(K)[i,j] = 1.\]
Let $W$ be the smallest power of $2$ larger than $\max_{ij}\{A[i,j]\}+\max_{ij}\{B[i,j]\}$. Then
$C(\frac{W}{2})$ gives the most significant bit of each entry in
$A \star B$. To obtain the second most significant bit, compute
$C(\frac{W}{4})$ and $C(\frac{3W}{4})$. The second bit of $(A
\star B)[i,j]$ is given by the expression:

\bigskip

\centerline{($\neg C(W)[i,j] \wedge C(\frac{3W}{4})[i,j])
\vee (\neg C(\frac{W}{2})[i,j] \wedge C(\frac{W}{4})[i,j])$.}

\bigskip

In general, to recover the first $k$ bits of $(A \star B)$,
one computes $C(\cdot)$ for $O(2^k)$ values of $K$. 
In particular, to obtain the $\ell$-th bits, compute

$$\bigvee_{s=0}^{2^{\ell-1}-1} [\neg C(W(1-\frac{s}{2^{\ell-1}}))\wedge C(W(1-\frac{s}{2^{\ell-1}}-\frac{1}{2^\ell}))].$$

\noindent To see this, notice that for a fixed $s$, if (for any $i,j\in [n]$)
$$W(1-\frac{s}{2^{\ell-1}}-\frac{1}{2^\ell})\leq \min_{k} A[i,k]+B[k,j]< W(1-\frac{s}{2^{\ell-1}}),$$
then the $\ell$-th bit of $\min_{k} A[i,k]+B[k,j]$ must be $1$, and if the $\ell$-th bit of $\min_{k} A[i,k]+B[k,j]$ is $1$, then there must exist an $s$ with the above property.

The values for $C(W(1-\frac{s}{2^{\ell-1}}))$ for even $s$ are needed for computing the $(\ell-1)$-st bits, hence to compute the $\ell$-th bits, at most $2^{\ell-2}+2^{\ell-1}$ dominance computations are necessary. To obtain the first $k$ bits of the distance product, one needs only $O(2^k)$ dominance product computations.
\end{proof}

\subsection{Buyer-Seller stable matching}

We show how the ``dominance-comparison''
ideas can be used to improve the runtime for solving a problem
arising in computational economics. In this problem, we have a set of
buyers and a set of sellers. Each buyer has a set of items he wants to purchase,
together with a maximum price for each item which he is willing to pay for that item.
In turn, each seller has a set of items she wishes to sell, together with a reserve price
for each item which she requires to be met in order for the sale to be completed. Formally:

\begin{definition}
An $(n,k)$-\emph{Buyer-Seller} instance consists of 
\begin{itemize}
\item a set $C=\{1,\ldots, k\}$ of \emph{commodities} \footnote{We will use
the words ``commodities'' and ``items'' interchangeably.}

\item an $n$-tuple of \emph{buyers} $B=\{b_1,\ldots, b_n\}$ where $b_i=(B_i,p_i)$, s.t. $B_i\subseteq C$ are the commodities desired by buyer $i$, and $p_i: B_i\rightarrow \R^{+}$ is the maximum price function for buyer $i$

\item an $n$-tuple of \emph{sellers} $S=\{s_1,\ldots, s_n\}$ where $s_i=(S_i,v_i)$, s.t. $S_i\subseteq C$ are the commodities owned by seller $i$, and $v_i: S_i\rightarrow \R^{+}$ is the reserve price function for seller $i$

\end{itemize}
\end{definition}

A sale transaction for an item $l$ between a seller who owns $l$ and a buyer who wants $l$ can take place if the 
price the buyer is willing to pay is at least the reserve price the seller has for the item. 
Let us imagine that each buyer wants to do business with only one
seller, and each seller wants to target a single buyer. Then the transaction between a buyer and a seller
 consists of all the items for which the buyer's maximum price meets the seller's reserve price.
 
\begin{definition}
Given a buyer $(B_i, p_i)$ and a seller $(S_j, v_j)$ the \emph{transaction set} $C_{ij}$ is defined as
$C_{ij}=\{l|\ l\in B_i\cap S_j,\ p_i(l)\geq v_j(l)\}$. Denote by $\mathbf{C}$ the \emph{transaction matrix} with entries $|C_{ij}|$.\\
The \emph{price} of $C_{ij}$ is defined as $P_{ij}=\sum_{l\in C_{ij}} p_i(l)$, and the \emph{reserve} of $C_{ij}$ is
 defined as $R_{ij}=\sum_{l\in C_{ij}} v_j(l)$. Denote by $P$ and $R$ respectively the transaction \emph{price} and \emph{reserve} matrices with entries $P_{ij}$ and $R_{ij}$.
\end{definition}

Further, we assume that every buyer $i$ has a preference relation on the sellers $j$ which depends entirely on $P_{ij}, R_{ij}$ and $|C_{ij}|$. Conversely, every seller has a preference relation on the buyers determined by the same three values. More formally, 

\begin{itemize}
\item buyer $i$ has a (computable) preference function $f_i: \R^{+}\times \R^{+}\times \Z^{+} \rightarrow \Z$ such that $i$ prefers seller $j$ to seller $j'$ iff $f_i(P_{ij},R_{ij},|C_{ij}|)\geq f_i(P_{ij'},R_{ij'},|C_{ij'}|)$. 

\item Similarly, seller $j$ has a (computable) preference function $g_j: \R^{+}\times \R^{+}\times \Z^{+} \rightarrow \Z$ such that $j$ prefers buyer $i$ to buyer $i'$ iff $g_j(P_{ij},R_{ij},|C_{ij}|)\geq g_j(P_{i'j},R_{i'j},|C_{i'j}|)$. 
\end{itemize}

Ideally, each buyer wants to talk to his most preferred seller, and each seller wants to sell to her most preferred buyer.
Unfortunately, this is not
always possible for all buyers, even when the prices and reserves
are all equal, and all preference functions equal $|C_{ij}|$. This is evidenced by the following example: Buyer
$1$ wants to buy item $2$, buyer $2$ wants to buy items $1$ and
$2$, seller $1$ has item $1$, seller $2$ has items $1$ and $2$.
Here buyer $1$ will not be able to get any items.

In a realistic setting, we want to find a buyer-seller matching so
that there is no pair $(b_i,s_j)$ for which $b_i$ is not paired with
$s_j$, such that both $b_i$ and $s_j$ would benefit from breaking their
matches and pairing among each other. This is the \emph{stable matching}
problem, for which optimal algorithms are known when the
preferences are known ({\em e.g.}, Gale-Shapley~\cite{GS62} can be
implemented to run in $O(n^2)$). However, for large $k$, the major
bottleneck in our setting is that of computing the preference
functions of the buyers and sellers.

The obvious approach to compute $P_{ij}, R_{ij}$ and $|C_{ij}|$ is to explicitly find the sets $C_{ij}$.
This gives an $O(k n^2)$ algorithm to compute $P_{ij}, R_{ij}$ and $|C_{ij}|$ for all pairs $(i,j)$.

Let for a (computable) function $f$, $T_f$ be the maximum time, over all $n$-bit $p,r$ and $c$, needed to compute $f(p,r,c)$. Let $T$ be the maximum time $T_f$, over all preference functions $f_i$ and $g_j$ for a buyer-seller instance. Then in time $O(kn^2 + Tn^2 +n^2\log n)$ one can obtain for every buyer (seller) a list of the sellers (buyers) sorted by the buyer's (seller's) preference function. 
Exploiting fast dominance computation, we can do better.

\begin{theorem}
The matrices $P$, $R$ and $\mathbf{C}$ for an $(n,k)$-Buyer-Seller instance can be determined in $O(n\sqrt{k M(n,k)})$ time, where $M(n,k)$ is the time required to multiply an $n\times k$ by a $k\times n$ matrix.
\end{theorem}

\begin{proof}
Using the dominance technique, we can compute matrix $\mathbf{C}$ as follows. 
For each buyer $i$ we create a $k$-dimensional vector
$\beta_i=\{\beta_{i1},\ldots,\beta_{ik}\}$ so that $\beta_{ij}=p_{i}(j)$ if $j\in
C_i$, and $\beta_{ij}=-\infty$ if $j\notin C_i$. For each seller $i$
we create a $k$-dimensional vector $\sigma_i=\{\sigma_{i1},\ldots, \sigma_{ik}\}$
so that $\sigma_{ij}=v_{i}(j)$ if $j\in S_i$, and $\sigma_{ij}=\infty$ if
$j\notin S_i$. Computing the dominance matrix for these points
computes exactly the number of items $l$ which buyer $i$ wants
to buy, seller $j$ wants to sell, and  $p_{i}(\ell)\geq v_{j}(\ell)$.

By a modification of Matousek's algorithm for computing
dominances, we can also compute the matrices $P$ and $R$.
We demonstrate how to find $R$. Recall that the dominance algorithm does a
matrix multiplication $A_k \cdot B_k^T$ with entries $A_k[i,j]=1$
iff $r_j(b_i)\in [ks, ks+s)$, and $B_k[i,j] = 1$ iff $r_j(s_i)
\geq ks+s$ (using the notation from Theorem~\ref{thm:matousek}). Let $B_k$ be the same, but redefine $A_k$ to be
\[A_k[i,j]=\begin{cases} v_{i}(j)& \textrm{if } r_j(b_i)\in [ks, ks+s)\\ 0 &
\text{otherwise}\end{cases}.\] Similar modifications are made to
the computation of the matrix $E$. Instead of adding $1$ to the
matrix entry $E[i_x,i]$ in the step for coordinate $j$, we add the corresponding reserve price $v_{i_x}(j)$. Determining $P$ can be done
analogously.
\end{proof}

\begin{corollary} A buyer-seller stable matching can be determined
in $O(n\sqrt{k M(n,k)} + n^2 \log n + n^2 T)$, where $T$ is the
maximum time to compute the preference functions of the
buyers/sellers, given the buyer price and seller reserve sums for
a buyer-seller pair.
\end{corollary}

For instance, if $k=n$ and $T=O(\textrm{polylog } n)$, the runtime of finding a
buyer-seller stable matching is
$O(n^{\frac{3+\omega}{2}})=O(n^{2.688})$.

\section{Heaviest $H$-subgraphs of real vertex-weighted graphs}

In the proof of Theorem \ref{t-vertex-real} it would be convenient to
assume that $H=K_h$ is a clique on $h$ vertices.
The proof for all other induced subgraphs with $h$ vertices is only slightly more cumbersome, but essentially the same.

Let $G=(V,E)$ be a graph with real vertex weights, and assume $V=\{1,\ldots,n\}$.
For two positive integers $a,b$, the {\em adjacency
system} $A(G,a,b)$ is the 0-1 matrix defined as follows.
Let $S_x$ be the set of all $\binom{n}{x}$ $x$-subsets of vertices.
The {\em weight} $w(U)$ of $U \in S_x$ is the sum of the weights of its elements. We {\em sort} the elements of $S_x$
according to their weights. This requires $O(n^x \log n)$ time, assuming $x$ is a constant.
Thus, $S_x=\{U_{x,1},\ldots,U_{x,\binom{n}{x}}\}$ where $w(U_{x,i}) \le w(U_{x,i+1})$.
The matrix $A(G,a,b)$ has its rows indexed by $S_a$. More precisely,
the $j$'th row is indexed by $U_{a,j}$. The columns are indexed by
$S_b$ where the $j$'th column is indexed by $U_{b,j}$.
We put $A(G,a,b)[U,U']=1$ if and only if $U \cup U'$ induces a $K_{a+b}$ in $G$
(this implies that $U \cap U' = \emptyset$), otherwise $A(G,a,b)[U,U']=0$.
Notice that the construction of $A(G,a,b)$ requires $O(n^{a+b})$ time.

For positive integers $a,b,c$, so that $a+b+c=h$, consider the
Boolean product $A(G,a,b,c)=A(G,a,b) \times A(G,b,c)$.
For $U \in S_a$ and $U' \in S_c$ for which $A(G,a,b,c)[U,U']=1$,
define their {\em maximum witness} $\delta(U,U')$ to be the maximal element $U'' \in S_b$
for which $A(G,a,b)[U,U'']=1$ and also $A(G,b,c)[U'',U']=1$.
For each $U \in S_a$ and $U' \in S_c$ with $A(G,a,b,c)[U,U']=1$
and with $U \cup U'$ inducing a $K_{a+c}$, if
$U''=\delta(U,U')$ then $U \cup U' \cup U''$ induces a $K_h$ in $G$ whose
weight is the largest of all the $K_h$ copies of $G$ that contain
$U \cup U'$. This follows from the fact that $S_b$ is sorted. Thus, by computing the maximum witnesses of all plausible pairs
$U \in S_a$ and $U' \in S_c$ we can find, for each pair of vertices, a $K_h$ in $G$ with maximum weight containing them, if such a $K_h$ exists,
or else determine that no $K_h$-subgraph contains the pair.

Let $A=A_{n_1 \times n_2}$ and $B=B_{n_2 \times n_3}$ be two 0-1 matrices.
The {\em maximum witness matrix} of $AB$ is the matrix $W=W_{n_1 \times n_3}$ defined as follows.
\begin{displaymath}
W[i,j] := \left\{\begin{array}{ll}
\textrm{the maximum } k~\textrm{s.t. } A[i,k]=B[k,j]=1 & \textrm{if } (AB)[i,j] \neq 0,\\
0 & \textrm{otherwise.}
\end{array}\right.
\end{displaymath}
Let $f(n_1,n_2,n_3)$ be the time required to compute the maximum witness
matrix of the product of an $n_1 \times n_2$ matrix by an $n_2 \times n_3$ matrix.
Let $h \ge 3$ be a fixed positive integer. For all possible choices of positive integers
$a,b,c$ with $a+b+c=h$ denote

$$
f(h,n) = \min_{a+b+c=h} f(n^a,n^b,n^c).
$$

Clearly, the time to sort $S_b$ and to construct $A(G,a,b)$ and $A(G,b,c)$ is overwhelmed
by $f(n^a,n^b,n^c)$. It follows from the above discussion that:
\begin{lemma}
\label{l21}
Let $h \ge 3$ be a fixed positive integer and let $G=(V,E)$ be a graph with $n$ vertices, each having a real weight.
For all pairs of vertices $u,v \in V$, an induced $H$-subgraph of $G$ containing $u$ and $v$ of maximum weight
(if one exists), can be found in
$O(f(h,n))$ time. Furthermore, if $f(n^a,n^b,n^c)=f(h,n)$ then the number of comparisons needed to find a maximum weight $K_h$ is $O(n^b \log n + z(G,a+c))$ where $z(G,a+c)$ is
the number of $K_{a+c}$ in $G$.
\end{lemma}
In fact, if $b \ge 2$, the number of comparisons in Lemma \ref{l21} can be reduced to only
$O(n^b+z(G,a+c))$. Sorting $S_b$ reduces to sorting the sums $X+X+\ldots+X$ ($X$ repeated $b$ times) of an $n$-element set of reals $X$. Fredman showed in \cite{Fr76a} that this can be achieved with only $O(n^b)$ comparisons.

A simple randomized algorithm for computing (not necessarily maximum) witnesses for
Boolean matrix multiplication, in essentially the same time required to perform the
product, is given by Seidel \cite{Se95}. His algorithm was derandomized by Alon and Naor
\cite{AlNa96}.
An alternative, somewhat slower deterministic algorithm was given by Galil and Margalit \cite{GaMa93}.
However, computing the matrix of maximum witnesses seems to be a more difficult problem. Improving an earlier algorithm of Bender et al. \cite{BeFaPeSkSu05},
Kowaluk and Lingas \cite{KoLi05} show that $f(3,n) = O(n^{2+1/(4-\omega)}) \le o(n^{2.616})$.
This already yields the case $h=3$ in Theorem \ref{t-vertex-real}.
We will need to extend and generalize the method from \cite{KoLi05} in order to obtain
upper bounds for $f(h,n)$. Our extension will enable us to answer more general queries such as ``is there a $K_h$ whose weight is within a given weight interval?''

{\bf Proof of Theorem \ref{t-vertex-real}:}\,
Let $h \ge 3$ be a fixed integer. Suppose $a,b,c$ are three positive integers
with $a+b+c=h$ and suppose that $0 < \mu \le b$ is a real parameter.
For two 0-1 matrices $A=A_{n^a \times n^b}$ and $B=B_{n^b \times n^c}$
the {\em $\mu$-split} of $A$ and $B$ is
obtained by splitting the columns of $A$ and the rows of $B$ into consecutive parts of size $\lceil n^\mu \rceil$ or $\lfloor n^\mu \rfloor$ each. In the sequel we ignore floors and ceilings whenever it does not affect the asymptotic nature of our results.
This defines a partition of $A$ into $p=n^{b-\mu}$ rectangular matrices $A_1,\ldots,A_p$,
each with $n^a$ rows and $n^\mu$ columns, and a partition of $B$ into $p$ rectangular matrices $B_1,\ldots,B_p$, each with $n^\mu$ rows and $n^c$ columns.
Let $C_i = A_iB_i$ for $i=1,\ldots,p$. Notice that each element of $C_i$ is a nonnegative integer of value at most $n^\mu$ and that $AB = \sum_{i=1}^p C_i$.
Given the $C_i$, the maximum witness matrix $W$ of the product $AB$
can be computed as follows. To determine $W[i,j]$ we look for the maximum index $r$
for which $C_r[i,j] \neq 0$. If no such $r$ exists, then $W[i,j]=0$; otherwise, having found $r$, we now look for the maximal index $k$ so that $A_r[i,k] = A_r[k,j]=1$. Having found $k$ we clearly have $W[i,j]=(r-1)n^\mu+k$.

We now determine a choice of parameters $a,b,c,\mu$ so that the time to compute
$C_1,\ldots,C_p$ and the time to compute the maximum witnesses matrix $W$,
is $O(n^{t(\omega,h)})$. By Lemma \ref{l21}, this suffices in order to prove the theorem.
We will only consider $\mu \le \min\{a,b,c\}$.
Taking larger values of $\mu$ results in worse running times.
The rectangular product $C_i$ can be computed by performing $O(n^{a-\mu}n^{c-\mu})$
products of square matrices of order $n^\mu$. Thus, the time required to compute $C_i$ is
$$
O(n^{a-\mu} n^{c-\mu} n^{\omega \mu}) = O(n^{a+c+(\omega-2)\mu}).
$$
Since there are $p$ such products, and since each of the $n^{a+c}$ witnesses can be
computed in $O(p+n^\mu)$ time, the overall running time is
$$
O(pn^{a+c+(\omega-2)\mu} + n^{a+c}(p+n^\mu))=O(n^{h-(3-\omega)\mu}+n^{h-\mu}+n^{h-b+\mu})
$$
\begin{equation}
\label{e6}
=O(n^{h-(3-\omega)\mu}+n^{h-b+\mu}).
\end{equation}
Optimizing on $\mu$ we get $\mu = b/(4-\omega)$. Thus, if, indeed,
$b/(4-\omega) \le \min\{a,c\}$ then the time needed to find $W$ is
$O(n^{h-b+b/(4-\omega)})$. Of course, we would like to take $b$ as large
as possible under these constraints.
Let, therefore, $b_1$ be the largest integer $b$ so that $b/(4-\omega) \le \lfloor (h-b)/2 \rfloor$. For such a $b_1$ we can take $a= \lfloor (h-b_1)/2 \rfloor$ and
$c = \lceil (h-b_1)/2 \rceil$ and, indeed, $\mu \le \min\{a,c\}$. Thus, (\ref{e6}) gives that the running time to compute $W$ is
$$
O(n^{h-b_1+b_1/(4-\omega)}).
$$
This justifies $s_1$ appearing in (\ref{e2}) in the definition of $t(\omega,h)$.
There may be cases where we can do better, whenever $b/(4-\omega) > \min\{a,c\}$.
We shall only consider the cases where
$a = \mu= \lfloor (h-b)/2 \rfloor \le b$ (other cases result in worse running times).
In this case $c=\lceil (h-b)/2 \rceil$ and, using (\ref{e6}), the running time is
$$
O(n^{h-(3-\omega)\lfloor \frac{h-b}{2} \rfloor}+n^{h-b+\lfloor \frac{h-b}{2} \rfloor}).
$$
This justifies $s_2$ appearing in (\ref{e4}) in the definition of $t(\omega,h)$.
Since $t(\omega,h)=\min\{s_1,s_2\}$ we have proved that
$W$ can be computed in $O(n^{t(\omega,h)})$ time. \npf

As can be seen from Lemma \ref{l21} and the remark following it, the number of comparisons that the algorithm performs is
relatively small. For example, in the case $h=3$ we have $a=b=c=1$ and hence the number of
comparisons is $O(n\log n+m)$. In all the three cases $h=4,5,6$ the value $b=2$ yields $t(\omega,h)$. Hence, the number of comparisons is $O(n^2)$ for $h=4$,
$O(n^2 + mn)$ for $h=5$ and $O(n^2 + m^2)$ for $h=6$.

Suppose $w: \{1,\ldots,n^b\} \rightarrow \Re$ so that $w(k) \le w(k+1)$.
The use of the $\mu$-split in the proof of Theorem \ref{t-vertex-real}
enables us to determine, for each $i,j$ and for a real interval $I(i,j)$, whether or not there exists an index $k$ so that $A[i,k]=B[k,j]=1$ and $w(k) \in I(i,j)$.
This is done by performing a binary search  within the $p=n^{b-\mu}$ matrices
$C_i, \ldots,C_p$. The running time in (\ref{e6}) only increases by a $\log n$ factor.
We therefore obtain the following corollary.
\begin{corollary}
\label{c21}
Let $H$ be a fixed graph with $h$ vertices, and let $I \subset \Re$.
If $G=(V,E)$ is a graph with $n$ vertices, and $w : V \rightarrow \Re$ is a weight function,
then, deciding whether $G$ contains an induced $H$-subgraph with total weight in $I$
can be done $O(n^{t(\omega,h)}\log n)$ time.
\end{corollary}

{\bf Proof of Theorem \ref{t-vertex-m}:}\,
We partition the vertex set $V$ into two parts $V = X \cup Y$ according to a parameter $\Delta$. The vertices in $X$ have degree at most $\Delta$. The vertices in $Y$ have degree
larger than $\Delta$. Notice that $|Y| < 2m/\Delta$.
In $O(m\Delta)$ time we can scan all triangles that contain a vertex
from $X$. In particular, we can find a heaviest triangle containing a vertex from $X$.
By Theorem \ref{t-vertex-real}, a heaviest triangle induced by $Y$ can be found in
$O((m/\Delta)^{t(\omega,3)})=O((m/\Delta)^{2+1/(4-\omega)})$ time.
Therefore, a heaviest triangle in $G$ can be found in
$$
O\left(m \Delta + \left(\frac{m}{\Delta}\right)^{2+1/(4-\omega)}\right)
$$
time. By choosing $\Delta=m^{(5-\omega)/(13-3\omega)}$ the result follows. \npf

The results in Theorems \ref{t-vertex-real} and \ref{t-vertex-m} are useful not only for real
vertex weights, but also when the weights are large integers. Consider, for example, the
graph parameter $\beta(G,H)$, the {\em $H$ edge-covering number} of $G$.
We define $\beta(G,H)=0$ if $G$ has no $H$-subgraph, otherwise $\beta(G,H)$ is the maximum
number of edges incident with an $H$-subgraph of $G$. 
To determine $\beta(G,K_k)$ we assign
to each vertex a weight equal to its degree. We now use the algorithm of Theorem \ref{t-vertex-real} to find the
heaviest $K_k$. If the weight of the heaviest $K_k$ is $w$, then $\beta(G,K_k)=w-\binom{k}{2}$.
In particular, $\beta(G,K_k)$ can be computed in $O(n^{t(\omega,k)})$ time.

Finally, we note that Theorems \ref{t-vertex-real} and \ref{t-vertex-m} apply also
when the weight of an $H$-subgraph is not necessarily defined as the sum of the weights of its vertices.
Suppose that the weight of a triangle $(x,y,z)$ is defined by a function
$f(x,y,z)$ that is monotone in each variable separately.
For example, we may consider $f(x,y,z)=xyz$, $f(x,y,z)=xy+xz+yz$ etc.
Assuming that $f(x,y,z)$ can be computed in constant time given $x,y,z$,
it is easy to modify Theorems \ref{t-vertex-real} and \ref{t-vertex-m} to find
a triangle whose weight is maximum with respect to $f$ in
$O(n^{2+1/(4-\omega)})$ time and $O(m^{(18-4\omega)/(13-3\omega)})$ time, respectively.

We conclude this section with the following proposition.
\begin{proposition}
\label{p21}
If $G=(V,E)$ is a graph with $n$ vertices, and $w : V \rightarrow \Re$ is a weight function,
then a (not necessarily induced) maximum weight $K_{2,k}$-subgraph can be found
in $O(n^{2+1/(4-\omega)})$.
\end{proposition}
\begin{proof}
To find the heaviest $K_{2,k}$ we simply need to find, for any two
vertices $i,j$, the $k$ largest weighted vertices $v_1,
\ldots, v_k$ so that each $v_i$ is a common neighbor of $i$ and
$j$. As in Lemma \ref{l21}, this reduces to finding the last $k$
maximum witnesses of a 0-1 matrix product. A simple modification
of the algorithm in Theorem \ref{t-vertex-real} achieves this goal
in the same running time (recall that $k$ is fixed).\end{proof}

\section{Heaviest $H$-subgraphs of real edge-weighted graphs}

Given a vertex-colored graph $G$ with $n$ vertices, an $H$-subgraph of $G$ is called {\em colorful} if each vertex of $H$ has a distinct color.
The {\em color coding} method presented in \cite{AlYuZw95} is based upon two important facts. The first one is that, in many cases, finding a colorful $H$-subgraph is easier than finding an $H$-subgraph in an uncolored graph. The second one is that in a random vertex coloring with $k$ colors, an $H$-subgraph with $k$ vertices becomes colorful with probability
$k!/k^k > e^{-k}$ and, furthermore, there is a derandomization technique that constructs
a family of not too many colorings, so that each $H$-subgraph is colorful in at least one of the colorings. The derandomization technique, described in \cite{AlYuZw95},
constructs a family of colorings of size $O(\log n)$ whenever $k$ is fixed.
It is based upon a
construction of {\em $k$-perfect hash functions} given in \cite{FKS84} and in \cite{ScSi90},
and constructions of small probability spaces that admit almost $\ell$-wise independent
random variables \cite{NaNa90}. The size of the constructed family of colorings is only
$O(\log n)$ where $k$ is fixed.

By the color coding method, in order to prove Theorem \ref{t-edge-cyc}, it suffices to
prove that, {\em given} a coloring of the vertices of the graph with $k$ colors, a colorful cycle of length $k$ of maximum weight (if one exists) can be found in $O(m^{2-1/\lceil k/2 \rceil})$ time.

{\bf Proof of Theorem \ref{t-edge-cyc}:}\,
Assume that the vertices of $G$ are colored with the colors $1,\ldots,k$.
We first show that for each vertex $u$, a maximum weight colorful cycle of length $k$ that passes through $u$ can be found in $O(m)$ time.
For a permutation $\pi$ of $1,\ldots,k$, we show that a maximum weight cycle
of the form $u=v_1,v_2,\ldots,v_k$ in which the color of $v_i$ is $\pi(i)$ can be found in
$O(m)$ time. Without loss of generality, assume $\pi$ is the identity.
For $j=2,\ldots,k$ let $V_j$ be the set of vertices whose color is $j$ so that
there is a path from $u$ to $v \in V_j$ colored consecutively by the colors $1,\ldots,j$.
Let $S(v)$ be the set of vertices of such a path with maximum possible weight.
Denote this weight by $w(v)$. Clearly, $V_j$ can be created from $V_{j-1}$ in $O(m)$ time
by examining the neighbors of each $v \in V_{j-1}$ colored with $j$.
Now, let $w_u = \min_{v \in v_k} w(v)+w(v,u)$. Thus, $w_u$ is the maximum weight of
a cycle passing through $u$, of the desired form, and a cycle with this weight can be retrieved as well.

We prove the theorem when $k$ is even. The odd case is similar. Let $\Delta=m^{2/k}$.
There are at most $2m/\Delta = O(m^{1-2/k})$ vertices with degree at least $\Delta$.
For each vertex $u$ with degree at least $\Delta$ we find a maximum weight colorful cycle of length $k$ that passes through $u$. This can be done in $O(m^{2-2/k})$ time.
It now suffices to find a maximum weight colorful cycle of length $k$ in the subgraph $G'$ of $G$ induced by the vertices with maximum degree less than $\Delta$.
Consider a permutation $\pi$ of $1,\ldots,k$. For a pair of vertices $x,y$, let
$S_1$ be the set of all paths of length $k/2$ colored consecutively by
$\pi(1), \ldots, \pi(k/2), \pi(k/2+1)$. There are at most $m\Delta^{k/2-1}=m^{2-2/k}$ such paths and they can be found using the greedy algorithm in $O(m^{2-2/k})$ time.
Similarly, let $S_2$ be the set of all paths of length $k/2$ colored consecutively by
$\pi(k/2+1), \ldots, \pi(k),\pi(1)$. If $u,v$ are endpoints of at least one path in $S_1$
then let $f_1(\{u,v\})$ be the maximum weight of such a path. Similarly define
$f_2(\{u,v\})$. We can therefore find, in $O(m^{2-2/k})$ a pair $u,v$ (if one exists)
so that $f_1(\{u,v\})+f_2(\{u,v\})$ is maximized. By performing this procedure for
each permutation, we find a maximum weight colorful cycle of length $k$ in $G'$. \npf

The definition of distance products, mentioned in the previous section, carries over to rectangular matrices.
Let $A=A_{n_1 \times n_2}$ and $B=B_{n_2 \times n_3}$ be two matrices with
entries in $\Re \cup \infty$.
In this case, the {\em distance product} $C = A \star B$ is an $n_1 \times n_3$ matrix
with $C[i,j] = \min_{k=1 \ldots ,n_2} A[i,k]+B[k,j]$.
By partitioning the matrices into blocks it is obvious that
Chan's algorithm for distance products of square matrices can be used to compute the distance product of an $n_1 \times n_2$ matrix
and an $n_2 \times n_3$ matrix in $O(n_1n_2n_3 /\log \min\{n_1,n_2,n_3\})$ time.
The analogous {\em MAX} version of distance products (namely, replacing MIN with MAX in the definition, and allowing $-\infty$) can be used to solve the MAX $H$-SUBGRAPH problem in edge weighted graphs.

{\bf Proof of Theorem \ref{t-edge-real}:}\,
We prove the theorem for $H=K_h$. The proof for other induced $H$-subgraphs is essentially the same. Partition $h$ into a sum of three positive integers
$a+b+c=h$. Let $S_a$ be the set of all $K_a$-subgraphs of $G$. Notice that
$|S_a| < n^a$ and that each
$U \in S_a$ is an $a$-set. Similarly define $S_b$ and $S_c$.
We define $A$ to be the matrix whose rows are indexed by $S_a$ and
whose columns are indexed by $S_b$. The entry $A[U,U']$ is defined to
be $-\infty$ if $U \cup U'$ does not induce a $K_{a+b}$, otherwise it is
defined to be the sum of the weights of the edges induced by $U \cup U'$.
We define $B$ to be the matrix whose rows are indexed by $S_b$ and
whose columns are indexed by $S_c$. The entry $A[U,U']$ is defined to
by $-\infty$ if $U \cup U'$ does not induce a $K_{b+c}$, otherwise, it is
defined to be the sum of the weights of the edges induced by $U \cup U'$
with {\em at least} one endpoint in $U'$.
Notice the difference in the definitions of $A$ and $B$.
Let $C = A \star B$. The time to compute $C$ using Chan's algorithm is
$O(n^h / \log n)$. Now, for each $U \in S_a$ and $U' \in S_c$ so that
$U \cup U'$ induces a $K_{a+c}$, let $w(U,U')$ be the sum of the
weights of the edges with one endpoint in $U$ and the other in $U'$ plus
the value of $C[U,U']$. If $w(U,U')$ is finite then it is the weight of
the heaviest $K_h$ that contains $U \cup U'$, otherwise no $K_h$ contains
$U \cup U'$. \npf

The weighted DENSE $k$-SUBGRAPH problem (see, e.g., \cite{FKP01})
is to find a $k$-vertex subgraph with maximum total edge weight. A
simple modification of the algorithm of Theorem \ref{t-edge-real}
solves this problem in $O(n^k/\log n)$ time. To our knowledge,
this is the first non-trivial algorithm for this problem. Note
that the maximum total weight of a $k$-subgraph can potentially be
much larger than a $k$-clique's total weight.

{\bf Proof of Theorem \ref{t-edge-cyc-2}:}\,
We use the color coding method, and an idea similar to Lemma 3.2 in \cite{AlYuZw95}. Given a coloring of the vertices with $k$ colors,
it suffices to show how to find the heaviest colorful path of length $k-1$ connecting any pair of vertices in $2^{O(k)}n^3/\log n$ time.
It will be convenient to assume that $k$ is a power of two,
and use recursion. Let $C_1$ be a set of $k/2$ distinct colors, and let
$C_2$ be the complementary set of colors. Let $V_i$ be the set of vertices
colored by colors from $C_i$ for $i=1,2$. Let $G_i$ be the subgraph induced by
$V_i$. Recursively find, for each pair of vertices in $G_i$, the maximum weight
colorful path of length $k/2-1$. We record this information in matrices
$A_1,A_2$, where the rows and columns of $A_i$ are indexed by $V_i$.
Let $B$ be the matrix whose rows are indexed by $V_1$ and whose columns are indexed by $V_2$ where $B[u,v]=w(u,v)$. The max-distance product
$D_{C_1,C_2}=(A_1 \star B) \star A_2$ gives, for each pair of vertices of $G$, all heaviest paths of length $k-1$ where the first $k/2$ vertices are colored by colors from $C_1$ and the last $k/2$ vertices are colored by colors from $C_2$.
By considering all $\binom{k}{{k/2}} < 2^k$ possible choices for $(C_1,C_2)$,
and computing $D_{C_1,C_2}$ for each choice, we can obtain an $n \times n$ matrix $D$ where $D[u,v]$ is the heaviest colorful path of length $k-1$ between $u$ and $v$. The number of distance products computed using this approach satisfies the recurrence $t(k) \le 2^k t(k/2)$. Thus, the overall running time
is $2^{O(k)}n^3/\log n$. \npf

The proof of Theorem \ref{t-edge-cyc-2} shows that, as long as
$k=o(\log \log n)$, a cycle with $k$ vertices and maximum weight can be
found, with high probability, in $o(n^3)$ time.
We note once again that all of our algorithms also apply to the {\em MIN} version of the problems.
The previous best known algorithm for finding a minimum weight cycle of length $k$, in real weighted graphs,
has running time $O(k!n^3 2^k)$ \cite{PlVo91}.

\section{Monochromatic and rainbow $H$-subgraphs}
{\bf Proof of Theorem \ref{t-rainbow}:}\,
Assume that $H$ has $t$ edges.
The problem of finding a rainbow $H$-subgraph in $G$ can be reduced, at
a small cost, to the problem of finding a rainbow $H$-subgraph in another edge-coloring
of $G$ where the number of colors used is only $t$.
Assume $C$ is the set of colors used in $G$ and consider a function $f: C \rightarrow \{1,\ldots,t\}$.
This defines a new edge-coloring of $G$. Clearly, if an $H$-subgraph is not rainbow in the
original coloring then it is also not rainbow in the new coloring.
If $f$ is constructed at random, a rainbow $H$-subgraph in the original coloring is also
rainbow in the new coloring with probability $t!/t^t > e^{-t}$. As in the color coding method, this method can be derandomized by constructing $O(\log m)=O(\log n)$ colorings
with only $t$ colors used in each of them, so that if $H$ is originally rainbow, it will also be rainbow in one of the constructed colorings.

We may now assume that $c: E \rightarrow \{1,\ldots,t\}$ and show how to find a rainbow $H$-subgraph if it exists, in $O(n^{\omega k +j})$ time.
We shall assume that $H=K_h$ and $h=3k+j$ where $j \in \{0,1,2\}$. The proof for other types of subgraphs is similar. By our assumption, $t=\binom{h}{2}$.
Consider a partition of $\{1,\ldots,t\}$ into 6 parts $C_1,C_2,C_3,C_4,C_5,C_6$.
The respective sizes are $|C_1|=|C_2|=\binom{k}{2}$, $|C_3|=\binom{k+j}{2}$,
$|C_4|=k^2$, $|C_5|=|C_6|=k(k+j)$. Notice that there are $2^{O(t)}$ choices for the partition. For each partition we construct two Boolean matrices $A$ and $B$ that are defined
as follows. The rows of $A$ are indexed by all the rainbow $K_k$ subgraphs of $G$ that
use the colors from $C_1$. The columns are indexed by all the rainbow $K_k$ subgraphs of $G$ that use the colors from $C_2$. We define $A[X,Y]=1$ if $X \cap Y = \emptyset$ and
the bipartite subgraph induced by the parts $X$ and $Y$ is complete, rainbow,
and uses the colors from $C_4$, otherwise $A[X,Y]=0$.
The rows of $B$ are indexed exactly in the same order as the columns of $A$.
Namely, by all the rainbow $K_k$ subgraphs of $G$ that use the colors from $C_2$. The columns are indexed by all the rainbow $K_{k+j}$ subgraphs of $G$ that use the colors from $C_3$. We define $B[X,Y]=1$ if $X \cap Y = \emptyset$ and
the bipartite subgraph induced by the parts $X$ and $Y$ is complete, rainbow,
and uses the colors from $C_5$, otherwise $A[X,Y]=0$.
The Boolean product $C=AB$ can be performed in $O(n^{\omega k+j})$ time using fast matrix multiplication. Now, if $C[X,Y] = 1$ and $X \cap Y = \emptyset$ and the bipartite subgraph of $G$ induced by the parts $X$ and $Y$ is complete, rainbow, and uses colors from $C_6$
then we must have that $X \cup Y$ is contained in a rainbow $K_h$ subgraph of $G$.
By considering all possible partitions we are guaranteed not to miss a single rainbow $K_h$-subgraph of $G$. \npf

{\bf Proof of Theorem \ref{t-monochromatic}:}\,
If $H$ is a star, the theorem is trivial. Next, assume that $H$ is not a star and has at least five vertices. Thus, $H$ has two independent edges and at least one additional vertex.
Put $h=3k+j$ and consider a labeling of the vertices of $H$ with $\{1,\ldots,h\}$
so that the following holds. If we partition $\{1,\ldots,h\}$ into three consecutive parts,
as equally as possible, then the subgraph of $H$ induced by the first part contains an edge
$e_1$ and the subgraph induced by the second part contains an edge $e_2$. Thus, e.g., if $H$ is the
$5$-cycle $(1,2,3,4,5)$ a plausible partition is $\{1,2\} \{3,4\} \{5\}$, $e_1=(1,2)$
and $e_2=(3,4)$.
Denote by $H_1$, $H_2$, and $H_3$, the labeled subgraphs of $H$ induced by each of the parts
and denote their respective sizes by $h_1,h_1,h_3$. Thus, if $j=0$ we must have
$h_1=h_2=h_3=k$, if $j=1$ we can assume $h_1=k+1$ and $h_2=h_3=k$ and if $j=2$ we can assume
$h_1=h_2=k+1$ and $h_3=k$.

We create a Boolean matrix $A$ as follows. The rows of $A$ are indexed by all the {\em ordered} $h_1$-tuples of vertices and the columns by all the {\em ordered} $h_2$-tuples.
We put $A[X,Y]=1$ if $X \cap Y=\emptyset$ and the mapping that assigns the $i$'th vertex of $X$ to $i$
and the $\ell$'th vertex of $Y$ to $h_1+\ell$ corresponds to a monochromatic labeled copy
of $H_1 \cup H_2$. In particular, note that the edge mapped to $e_1$ has the same color as the edge mapped to $e_2$. We create a Boolean matrix $B$ as follows. The rows of $B$ are indexed
exactly like the columns of $A$. The columns of $B$ are indexed by all the ordered $h_3$-tuples.
We put $B[X,Y]=1$ if $X \cap Y = \emptyset$ and the mapping that assigns the $i$'th vertex of $X$ to $h_1+i$
and the $\ell$'th vertex of $Y$ to $h_1+h_2+\ell$ corresponds to a monochromatic labeled copy
of $H_2 \cup H_3$. Let $C=AB$ be the Boolean product. Suppose that $C[X,Y]=1$ and suppose
also that $X \cup Y$ corresponds to a monochromatic labeled copy of $H_1 \cup H_3$.
Let $Z$ satisfy $A[X,Z]=1$ and $B[Z,Y]=1$. Then we must have that $X \cup Y \cup Z$ corresponds to a monochromatic copy of $H$. This is because the color of each mapped edge is either that of the edge of $G$ mapped to $e_1$ or that of the edge of $G$ mapped to $e_2$ but these two are also colored the same. Notice also that if there is a monochromatic $H$-subgraph, it would be captured by our algorithm. Since the time needed to compute $C$ is
$O(n^{\omega k +j})$, the result follows.

Consider next the case $h=4$. If $H \neq K_4$ then we can assume that $H$ is labeled by $\{1,2,3,4\}$ so that $(1,4)$ is {\em not} an edge and $(2,3)$ is an edge.
Thus, the same algorithm described above using the partition $\{1\} \{2,3\} \{4\}$ yields
an $O(n^{\omega+1})$ time algorithm for detecting a monochromatic $H$. If $H=K_4$ the algorithm is slightly different. For each $v \in G$, let $S_1,\ldots,S_t$ be a partition of the neighbors of $v$ so that $x,y \in S_i$ if and only if $c(v,x)=c(v,y)$.
Searching for a triangle in the subgraph induced by $S_i$ all of whose edges are colored
by a given specific color has the same complexity as searching for an uncolored triangle
in a graph, and hence can be done in $O(|S_i|^\omega)$ time. Thus, in
$O(\sum_{i=1}^t |S_i|^\omega) \le O(n^\omega)$ we can find a monochromatic triangle containing $v$, if it exists. Performing this procedure for each $v \in V$ gives the desired $O(n^{\omega+1})$ time algorithm.

The only remaining case is $H=K_3$. Let $E_i$ be the set of edges of $G$ colored with $i$.
We say that $i$ is {\em heavily used} if $|E_i| \ge n^{(\omega+1)/2}$.
Clearly, the number of colors heavily used is at most $O(n^{2-(\omega+1)/2})$. For each heavily used color $i$ we can decide, in $O(n^\omega)$ time, whether there is a monochromatic triangle colored with $i$. The overall running time is, therefore
$O(n^{\omega+2-(\omega+1)/2}) = O(n^{(3+\omega)/2})$.
For each color $i$ that is not heavily used, we can decide in $O(|E_i|^{2\omega/(\omega+1)})$ time whether there is a monochromatic triangle colored with $i$ using the algorithm from \cite{AlYuZw97}.
The overall running time is maximized if $|E_i|=\Theta(n^{(\omega+1)/2})$ and when there are
$\Theta(n^{2-(\omega+1)/2})$ such colors. In this extremal case the running time is still
only $O(n^{((\omega+1)/2)(2\omega/(\omega+1))+2-(\omega+1)/2})=O(n^{(3+\omega)/2})$. \npf

\section{Concluding remarks and open problems}

We presented several algorithms for MAX $H$-SUBGRAPH in both real vertex weighted or real edge weighted graphs, and results for the related problem of finding
monochromatic or rainbow $H$-subgraphs in edge-colored graphs.
It may be possible to improve upon the running times of some of our algorithms.
More specifically, we raise the following open problems.\\
(i) Can the exponent $t(\omega,3)$ in Theorem \ref{t-vertex-real} be improved?
If so, this would immediately imply an improved algorithm for maximum witnesses.\\
(ii) Can the logarithmic factor in Theorem \ref{t-edge-cyc} be eliminated?
We know from \cite{AlYuZw97} that this is the case in the unweighted version of the problem. Can the logarithmic factor in Theorem \ref{t-rainbow} be eliminated?\\
(iii) Can monochromatic triangles be detected faster than the
$O(n^{(3+\omega)/2})$ algorithm of Theorem \ref{t-monochromatic}?
In particular, can they be detected in $O(n^\omega)$ time?\\
(iv) 
Is the dominance matrix for a set of $n$ points in $n$ dimensions computable in $\tilde{O}(n^{\omega})$ time?

\section*{Acknowledgment}
The authors thank Uri Zwick for some useful comments.

\end{document}